\pdfoutput=1

\documentclass[twocolumn,twoside,slac_two]{revtex4}
\usepackage{graphicx}
\usepackage{fancyhdr}
\usepackage[raggedright]{titlesec}
\usepackage{amsmath}
\usepackage{amssymb}
\usepackage{xspace}
\usepackage{nicefrac}

\renewcommand{\thesection}{\arabic{section}}
\renewcommand{\thesubsection}{\arabic{subsection}}
\titleformat{\section}
  {\bfseries\large}{\thesection.}{0.5em}{}
\titleformat{\subsection}
  {\bfseries}{\thesection.\thesubsection.}{0.5em}{}
\titleformat{\paragraph}[runin]
  {\bfseries}{\theparagraph.}{0.5em}{}[\phantom{}]

\titlespacing{\section}{0pt}{24pc plus 24pc minus 0.4pc}{6pt plus .3pt}
\titlespacing{\subsection}{0pt}{8pt plus 2pt minus 0.3pt}{3pt plus .3pt}
\titlespacing{\section}{0pt}{*1.5}{*0.75}
\titlespacing{\subsection}{0pt}{*1}{*0.5}

\newcommand{\nue}{\ensuremath{\nu_e}\xspace}
\newcommand{\nuebar}{\ensuremath{\overline{\nu}\hspace{-0.5pt}_e}\xspace}
\newcommand{\numu}{\ensuremath{\nu_\mu}\xspace}
\newcommand{\nutau}{\ensuremath{\nu_\tau}\xspace}
\newcommand{\nui}{\ensuremath{\nu_1}\xspace}
\newcommand{\nuii}{\ensuremath{\nu_2}\xspace}
\newcommand{\nuiii}{\ensuremath{\nu_3}\xspace}

\newcommand{\dmsq}[1]{\ensuremath{\Delta m^2_{#1}}\xspace}

\newcommand{\ue}[1]{\ensuremath{\mathrm{e}^{#1}}\xspace}
\newcommand{\ui}{\ensuremath{\mathrm{i}}\xspace}
\newcommand{\rtfr}[2]{\ensuremath{\sqrt{\nicefrac{#1}{#2}}}\xspace}

\newcommand{\braket}[2]{\ensuremath{\left\langle{#1\vphantom{#2}}\right|\left.\!{\vphantom{#1}#2}\right\rangle}\xspace}

\newcommand{\abs}[1]{\ensuremath{\left|#1\right|}\xspace}
\newcommand{\modsq}[1]{\ensuremath{\abs{#1}^2}\xspace}

\newcommand{\unit}[1]{\ensuremath{\,\mathrm{#1}}\xspace}

\newcommand{\fig}[1]{FIG~\ref{#1}\xspace}
\renewcommand{\prl}{Phys.\,Rev.\,Lett.\,}
\begin{document}

\title{(Direct) Measurement of $\mathbf{\theta_{13}}$}

\author{R.\,P.\,Litchfield}
\affiliation{Department of Physics, Kyoto University, Kyoto, Japan / \\ Department of Physics, University of Warwick, Coventry, United Kingdom}

\begin{abstract}
A review of recent measurements of the neutrino mixing angle $\theta_{13}$ is presented.  In the standard parametrisation of three-neutrino mixing this is the last of the three mixing angles to be determined, and is known to be the smallest.  The angle parametrises the overlap \modsq{\braket{\nuiii}{\nue}}, a non-zero value of which is necessary for leptonic CP violation driven by the KM mechanism.  A non-zero overlap also makes experimental determination of the neutrino mass hierarchy much easier.  Several experiments have recently reported non-zero measurements of $\theta_{13}$; this presentation concentrates on T2K, MINOS, RENO and Double Chooz. 
\end{abstract}

\maketitle

\thispagestyle{fancy}

\section{Importance of $\mathbf{\theta_{13}}$}
The phenomenon of neutrino oscillations, in the form of disappearance experiments in `atmospheric' and `solar' $L/E$ ranges is well established.  In the standard 3-neutrino picture these constrain 2 of the (3 real, 1 complex phase) parameters that specify the $3\times3$ PMNS neutrino mixing matrix.  It has been known for around a decade that the PMNS matrix is very different from the quark-sector CKM matrix, in that most elements are large.  The (in general complex) elements $U_{\alpha i} = \braket{\nu_i}{\nu_\alpha}$ correspond to the overlaps between mass and flavour eigenstates, and to a first approximation, the matrix has tribimaximal form~\cite{tbm}
\begin{equation}
\begin{pmatrix}
    U_{e1} & U_{e2} & U_{e3} \\
    U_{\mu1} & U_{\mu2} & U_{\mu3} \\
    U_{\tau1} & U_{\tau2} & U_{\tau3}
  \end{pmatrix} \simeq \begin{pmatrix}
    \rtfr{2}{3} & \rtfr{1}{3} & 0 \\
    \rtfr{1}{6} & \rtfr{1}{3} & \rtfr{1}{2} \\
    \rtfr{1}{6} & \rtfr{1}{3} & \rtfr{1}{2} \\
  \end{pmatrix}\textrm{.}
\end{equation}

The absolute mass of the neutrino mass eigenstates is unknown, but the splittings $\dmsq{ji} = m^2_j - m^2_i$ that drive the oscillations are mostly known. We do not yet know if \nuiii is the heaviest or lightest of the neutrino mass eigenstates, situations referred to as the normal and inverse hierarchy, respectively.  
\subsection{Angular parametrisation}
In the standard parametrisation, the  mixing matrix is decomposed as a product of three transformations
\begin{equation}
U = R(\theta_{12})U(\theta_{13}, \delta)R(\theta_{23})\textrm{,} 
\end{equation}
where $R(\theta_{12})$ and $R(\theta_{23})$ are the Givens rotations between \mbox{\nui--\nuii}  and \mbox{\numu--\nutau} respectively, and $U(\theta_{13}, \delta)$ is the unitary transform with real leading diagonal between \mbox{\nue--\nuiii}. This parametrisation is convenient because to a first approximation the most studied channels (solar-scale \nue disappearance and atmospheric-scale \numu disappearance) provide direct measurements of the two rotation matrices.  In this parametrisation, if $\theta_{13}$ was zero the unitary sub-matrix would be diagonal. Thus the PMNS matrix would be purely real and there would be no (KM-mechanism) CP violation in the neutrino sector.  Multiplying out the sub-matrices we find that $\theta_{13}$ is most closely related to the element $U_{e3}$, to which it is related by
\begin{equation}
U_{e3} = \sin\theta_{13}\ue{-\ui\delta} \ \Leftrightarrow \ \sin^2\theta_{13} = \modsq{U_{e3}}\textrm{.}
\end{equation} 
Indeed this definition provides our best physical interpretation of $\theta_{13}$: it describes the magnitude of overlap between \nuiii and \nue.  Not surprisingly, the easiest measurements of $\theta_{13}$ are those involving the element $U_{e3}$.  With available neutrino sources, the most accessible channels are $\numu\rightarrow\nue$ and $\nuebar\rightarrow\nuebar$, at the first atmospheric maximum (corresponding to  $L/\mathrm{m} \sim 500 \times E/\mathrm{MeV}$).

\section{Long-baseline experiments}
Here we refer to experiments done with accelerator-produced beams of neutrinos, mainly \numu, with baselines longer than 100\unit{km}.  

Accelerator experiments use the $\numu \rightarrow \nue$ oscillation channel, and search for appearance of \nue (above backgrounds from beam contamination) in a distant 'far' detector.  A good pedagogical~\cite{Freund}
 expression for the appearance probability expands it in terms of small parameters, $\sin2\theta_{13}$ and $\alpha = \dmsq{21}/\dmsq{31} \simeq 1/32\!$ , and retains only the quadratic terms:
\begin{equation}
\begin{split}
&P(\numu\rightarrow\nue) \simeq\\ 
  &\ \phantom{+} T_{\theta\theta}  \sin^2(2\theta_{13}) \frac{\sin^2\left([1-A]\Delta\right)}{[1-A]^2}\\
  &\ + T_{\alpha\alpha}  \alpha^2 \frac{\sin^2(A\Delta)}{A^2}\\
  &\ + T_{\alpha\theta}  \alpha \sin(2\theta_{13}) \frac{\sin\left([1-A]\Delta\right)}{1-A}\frac{\sin(A\Delta)}{A}\cos(\delta + \Delta)	\textrm{,}
\end{split}
\end{equation}
where
\begin{flalign*}
\Delta &= \frac{\dmsq{31} L}{4E} & \text{$\sim \frac{\pi}{2}$ at $1^\mathrm{st}$  oscillation maximum}\\
A &= \pm\frac{2\sqrt{2}G_{F}n_{e}E}{\abs{\dmsq{31}}} & \text{is the matter density parameter~\footnotemark[9],}
\end{flalign*}
\vphantom{\footnote[9]{for MINOS $\abs{A}\sim0.3$, for T2K $\abs{A}\sim0.07$}}
and the $T$ coefficients are
\begin{align*}
T_{\theta\theta}^{\hphantom{\alpha\alpha}} &=\sin^2\theta_{23}\\
T_{\alpha\alpha}^{\hphantom{\alpha\alpha}} &= \cos^2\theta_{23}\sin^2(2\theta_{12})\\
T_{\alpha\theta}^{\hphantom{\alpha\alpha}} &= \cos\theta_{13}\sin(2\theta_{12})\sin(2\theta_{23})\textrm{.}
\end{align*}
There are two experiments of this kind that have recently produced results: MINOS~\cite{MINOS}  and T2K~\cite{T2K}. MINOS is an older experiment, and was optimised for studying \numu disappearance, not \nue appearance. But it is a mature experiment and currently has the largest data set of any long-baseline experiment. T2K is newer and will be much more sensitive, but currently has only a small fraction of its planned integrated luminosity.  Both experiments follow the same basic principle:
\begin{itemize}
\item A beam of \numu from pion decays.
\item A near detector to measure interaction rates.
\item A far detector to look for oscillations.
\end{itemize}
\subsection{Muon neutrino beams}
Both MINOS and T2K use similar beamlines to produce muon neutrinos.  Each beamline consists of a graphite target placed in a proton beam in order to produce pions.  The target sits inside the first of two (MINOS) or three (T2K) magnetic horns, which focus charged secondary particles of a given charge sign (more commonly positive secondaries).  The geometry of the horns and decay volume with respect to the target determines the momentum range for which secondaries will be focussed. 

The focussed secondary beam is mostly pions, and these are allowed to decay in flight in a helium-filled decay pipe to produce (predominantly) $\mu^+$ and \numu.  Downstream of the decay volume, instrumentation allows characterisation of the beam by looking at the profile of muons and remaining hadrons.  The beams are angled a few degrees downward due to the curvature of the earth, and after passing through the ground for around 100m only neutrinos remain. The contamination from antineutrinos and \nue is typically a few percent.

In T2K an additional trick is used.  The detectors are placed slightly off of the beam axis ($\sim2.5^\circ$ at the target), and the on-axis neutrino beam energy is tuned a little way above the desired energy. At a small angle to the parent direction the neutrino energy as a function of parent momentum turns over in the lab frame, resulting in a neutrino spectrum that is insensitive to the parent energy.  This gives a slightly narrower flux peak compared to an on-axis beam tuned for the same energy, and drastically reduces the tail out at higher energies. Both of these effects are useful for \nue appearance experiments; the latter one particularly so, as feed down from neutral current interactions of high energy neutrinos is a potentially large background to the \nue charge current interaction signal.


\subsection{MINOS analysis}
Both MINOS detectors are comprised of steel/scintillator tracking calorimeters.  At a typical energy of 3\unit{GeV} the \nue signal events appear as small tight showers, and a multi-variate discriminant, based on the matching to a large library of template events, is used to separate these from the (generally more diffuse) hadronic showers from neutral current backgrounds.  A large background remains however, and the analysis make much use of similarity of the near and far detectors to constrain the background contributions.  The analysis is also significantly helped by the fact that MINOS's beam geometry is adjustable, which provides an mechanism to separate flux and cross-section uncertainties.  The background is therefore well-controlled, but its size limits the sensitivity of the analysis.

\subsection{T2K analysis}
T2K makes use of the 22.5\unit{kt} (fiducial) Super-Kamiokande water-\v{C}hernkov detector as its far detector.  The flux peak is in the hundred-MeV region and at these energies the \nue signal is from the charge current quasi-elastic interaction, $\nue+n\rightarrow p+ e^-$ on $\vphantom{O}^{16}\textrm{O}$ nuclei, which is detected  as a fuzzy \v{C}erenkov ring.  Quasi-elastic kinematics allows the neutrino energy to be computed from the reconstructed energy and direction of the ring, while discrimination from the more common $\mu^-$ events is based on the fuzziness of the ring. The remaining background is dominated by two sources, the first of which is contamination from \nue that are in the original, unoscillated, beam.  These predominantly come from kaon decays, and in the off-axis geometry, typically have higher energy than the oscillation signal. They can be reduced with a cut on reconstructed energy.   The second major background is from neutral current events that produce a single energetic $\pi^0$.  The photon pair from the $\pi^0$ decay can imitate an electron shower if one ring is not resolved.  This kind of background can be reduced by specifically looking for a secondary ring.  

The importance of exclusive interaction channels, both as signal and sources of background mean that T2K needs a good understanding of cross-sections; this drives the design of the near detectors.  The near detector complex has detector sub-systems both on- and off-axis. The off-axis suite uses fine-resolution scintillator detectors as a neutrino target regions, and time-projection chambers for momentum measurement and particle identification.  This can be used directly to provide a much better estimate of interaction rates at Super-Kamiokande than from simulation alone.  There is also an extensive program of dedicated cross-section measurements to improve neutrino interaction simulations in the long term.  The on-axis INGRID detector has a rather different primary role.  Because of the effect of off-axis angle on the neutrino spectrum, the neutrino beam direction must be stable to high precision (1\unit{mrad}).  INGRID is a cross-shaped arrangement of iron/scintillator modules, and provides a high event rate and large lever arm to monitor the neutrino beam direction.  

T2K is not able to tune its beam to the same extent as MINOS, so the flux predictions are more important. To this end, a replica target was provided to and used in the NA61 hadroproduction experiment at CERN.  The primary beam energy at NA61 is also very close to that used by T2K meaning there is very little extrapolation to be made from NA61 data to T2K flux inputs.

\subsection{Results from appearance analyses}
\begin{figure}
\centering
\includegraphics[width=0.45\columnwidth]{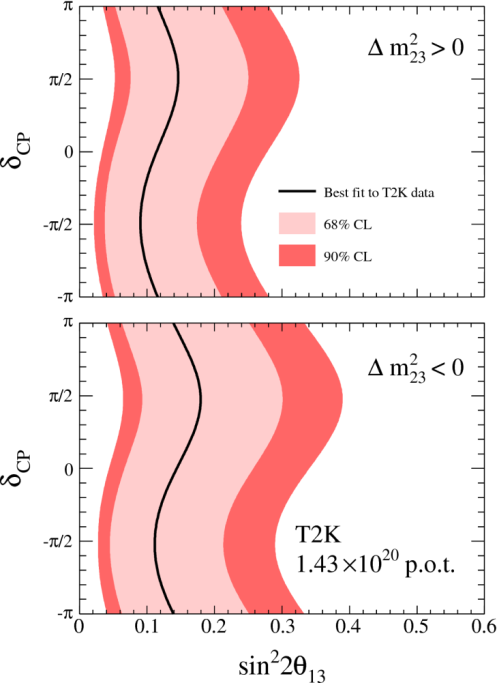}
\includegraphics[width=0.43\columnwidth]{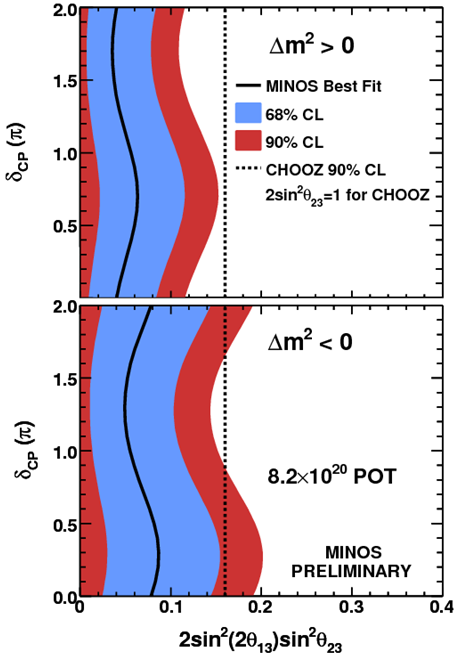}
\caption{For T2K (L) and MINOS (R): measured value of $\sin^22\theta_{13}$ for different assumed values of $\delta$.  Note that the $y$-axes are offset by $\pi$, and that treatment of $\theta_{12}$, $\theta_{23}$ and $\dmsq{}$  is different: T2K holds these fixed, whilst MINOS allows them to vary within global uncertainties.}
\label{lbne-q13}
\end{figure}

The results from MINOS and T2K have comparable sensitivity at this stage, and are shown in \fig{lbne-q13}.  Characteristic of $\numu\rightarrow\nue$ experiments is a correlation between $\sin^22\theta_{13}$ and the CP phase $\delta$, due to the $T_{\alpha\theta}$ term in the probability.  Both experiments favour a finite value of $\sin^22\theta_{13}$, with MINOS and T2K disfavouring $\sin^22\theta_{13}=0$ at 89\%, and 99.3\% respectively.  The different confidence levels primarily reflect the different central values. 

\section{Reactor experiments} 
Here we consider the current generation of reactor experiments in a generic way, as they all operate in a very similar way.  The Daya Bay experiment is covered in more detail in~\cite{DBatFPCP}. For details on the individual experimental results see~\cite{DC, DB, RENO}. 

\subsection{Anti-neutrino disappearance}
Reactor experiments look for the disappearance of \nuebar in the flux from the operating fission reactors. This provides a cheap, intense source of neutrinos, in the energy range of a few MeV.  The signal channel is the inverse beta decay reaction ($\nuebar + p \rightarrow n+ e^+$).

The interaction rate is the product of the falling reactor flux and the rising inverse beta decay cross-section, and peaks in the range of $3\sim4\unit{MeV}$, with a tail out to about 10\unit{MeV}. For neutrinos in this energy range we can ignore matter effects in the earth's crust and the survival probability is quite simple:
\begin{equation}
\begin{split}
P(\nuebar\rightarrow\nuebar) \simeq 1 &- \sin^2(2\theta_{13})\sin^2\Delta  \\
	&- \cos^4\theta_{13}\sin^2(2\theta_{12})\sin^2(\alpha\Delta)\textrm{.}
\end{split}
\end{equation} 
On a 1\unit{km} baseline, the third (solar scale) term is around 0.001, around 1\% of the second (atmospheric scale) term, so it contributes very little uncertainty to the probability.  As a result the survival probability gives direct access to $\sin^2(2\theta_{13})$.  In contrast with the long baseline experiments, measurement of $\theta_{13}$ with reactor experiments is theoretically clean but, by the same token, they cannot determine the mass hierarchy or look for CP violation. 

Previous anti-neutrino experiments using reactor sources include the Savannah River experiment that first detected the neutrino, the CHOOZ experiment that gave the previous best measurement of $\sin^2(2\theta_{13})$, and the KamLAND experiment which, using a longer baseline, measured \dmsq{21}  and $\sin^2(2\theta_{12})$.

\subsection{Current-generation reactor experiments}
The three new experiments are all quite similar in terms of basic principles.  For definiteness numbers from RENO are used unless specified, but all experiments are similar, with Daya Bay typically being somewhat larger, and Double Chooz slightly smaller.

The most important improvement over previous experiments is a trick taken from the long baseline experiments: the use of a near detector to eliminate uncertainties in the reactor flux.  The near and far baselines are not as different as in the long baseline experiments however, so the experiments can use essentially identical detectors at their near and far sites. As an example, the RENO near hall is on a weighted baseline of 409\unit{m} and the far baseline only 3 times more distant at 1444\unit{m}. Because of this, oscillations at the near site must also be taken into account when analysing the data.

Another very important consideration is the event rate. To this end all experiments are built on the site of large, multi-core, power plants.  Because of the multiple cores and detector locations, care must be taken to sum contributions correctly.  For example, the fluxes from each reactor must be weighted by live time, and the distance from the reactor core to the detector. The individual distances must also be used when calculating the survival probability. 

\begin{figure}
\centering
\includegraphics[width=0.95\columnwidth]{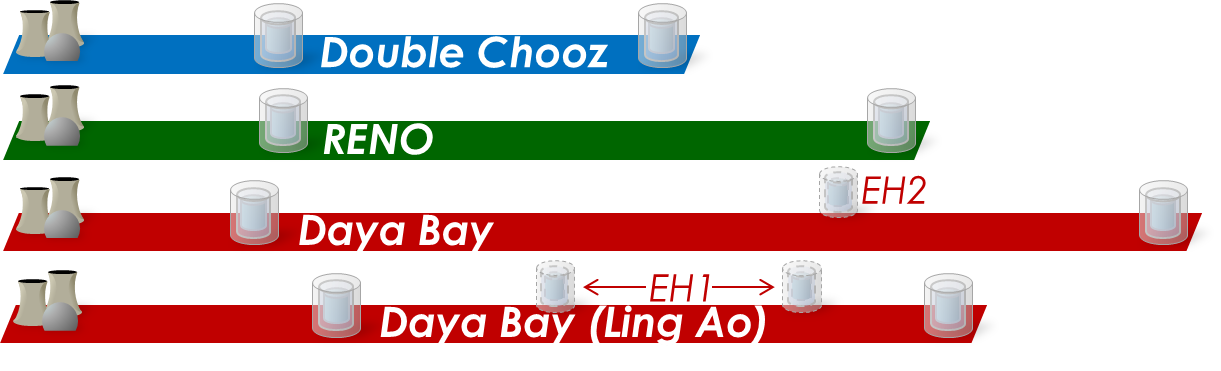}
\caption{Cartoon of baselines (weighted average over reactor cores) for reactor experiments. In Daya Bay there is an additional contribution because the Daya Bay near hall EH1 is at an intermediate distance from the two Ling Ao reactor sites, and vice versa for EH2.}
\label{baselines}
\end{figure}

A comparison of the baselines is shown in \fig{baselines}. Double Chooz is the simplest case with two 4.3\unit{GW} cores and the near detector placed close to the same flux-ratio locus as the far detector.  RENO has a more complicated layout, with six 2.8\unit{GW} cores on a perpendicular line from the near--far axis.  Daya Bay's layout is the most complicated of all, with six 2.9\unit{GW} cores at three reactor sites, and three experimental halls.

The three experiments all use the same basic technology for antineutrino detection. To reduce backgrounds the experiments use a delayed coincidence technique.  The target volume is a roughly $2\sim3\unit{m}$ clear acrylic vessel containing liquid scintillator doped with gadolinium.  The prompt signal comes from the positron scintillation light plus the eventual annihilation photons. The delayed coincidence comes from neutron capture on the gadolinium.  Prompt and delayed signal photons are detected by inward facing PMTs, but between them and the target there are two more detector regions.   Immediately surrounding the target vessel is a `gamma-catcher' region which is around 0.5\unit{m} thick and contains liquid scintillator without gadolinium dopant.  This improves energy reconstruction, and because the gamma catcher contains no gadolinium, the fiducial volume is defined physically, avoiding any systematic error resulting from position reconstruction.  The gamma catcher itself is inside a clear vessel, and is surrounded by a buffer region (mineral oil without scintillator), up to 1\unit{m} thick, which is designed to insulate the active detectors from background events originating from the PMTs, which are mounted on the outer wall of the buffer region.  Outside this there is typically a water-\v{C}hernkov veto tank, used for detecting external radioactivity backgrounds and cosmic muons that can induce backgrounds from following neutrons.

Double Chooz uses two such detectors, one at the near site and one at the far site, but the current results pre-date the commissioning of the near detector, and estimate the far detector event rate directly from simulations.  RENO also uses two similar detectors, and the first analysis already includes data from both detectors.  The larger Daya Bay experiment has six operating detectors, two at the Daya Bay near hall (EH1), one at the Ling Ao near hall (EH2) and three at the far hall (EH3).  A further two detectors are still to be installed, one each in EH2 and EH3.

\begin{figure}
\centering
\includegraphics[width = 0.48\columnwidth]{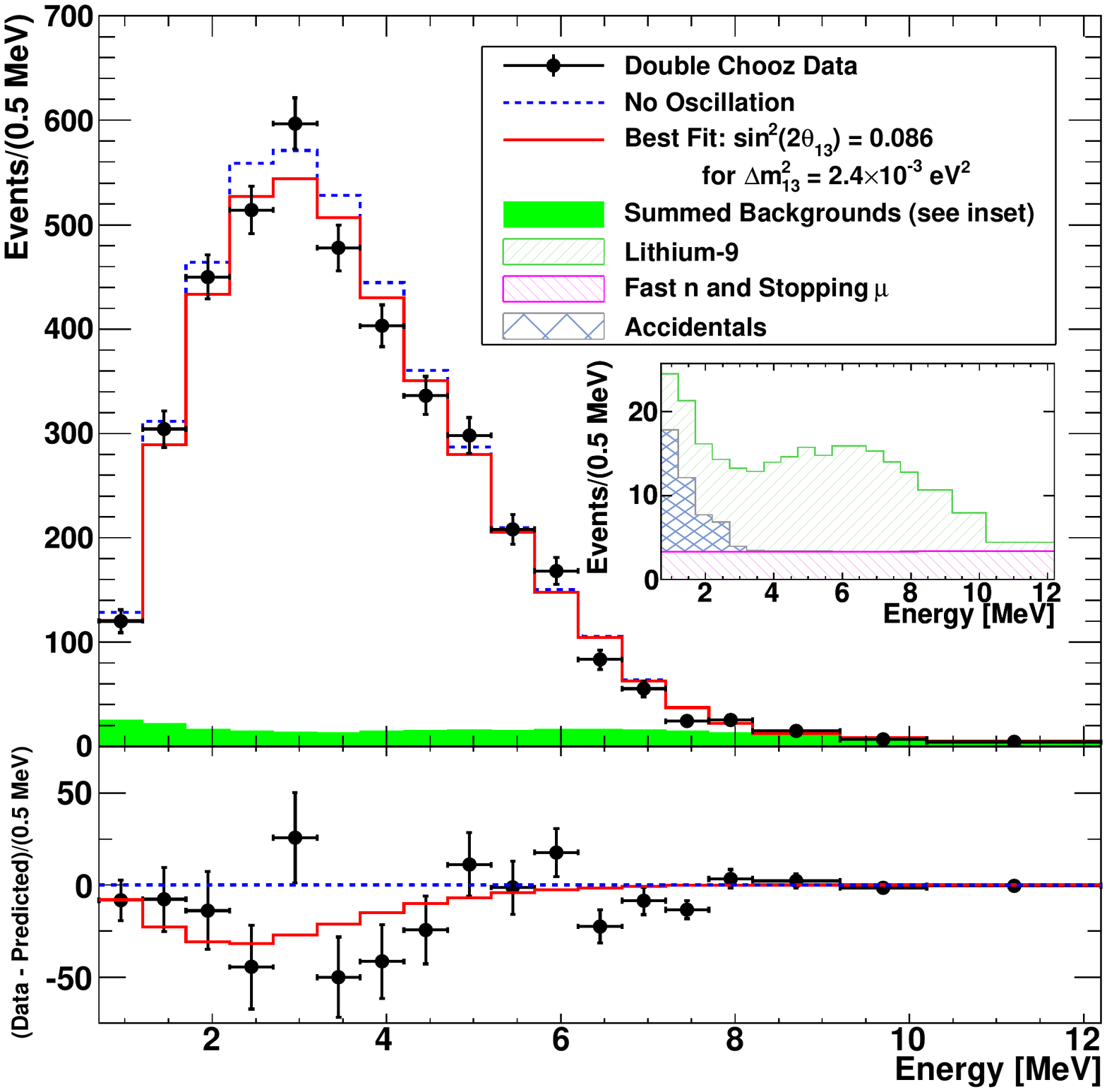}
\includegraphics[width = 0.48\columnwidth]{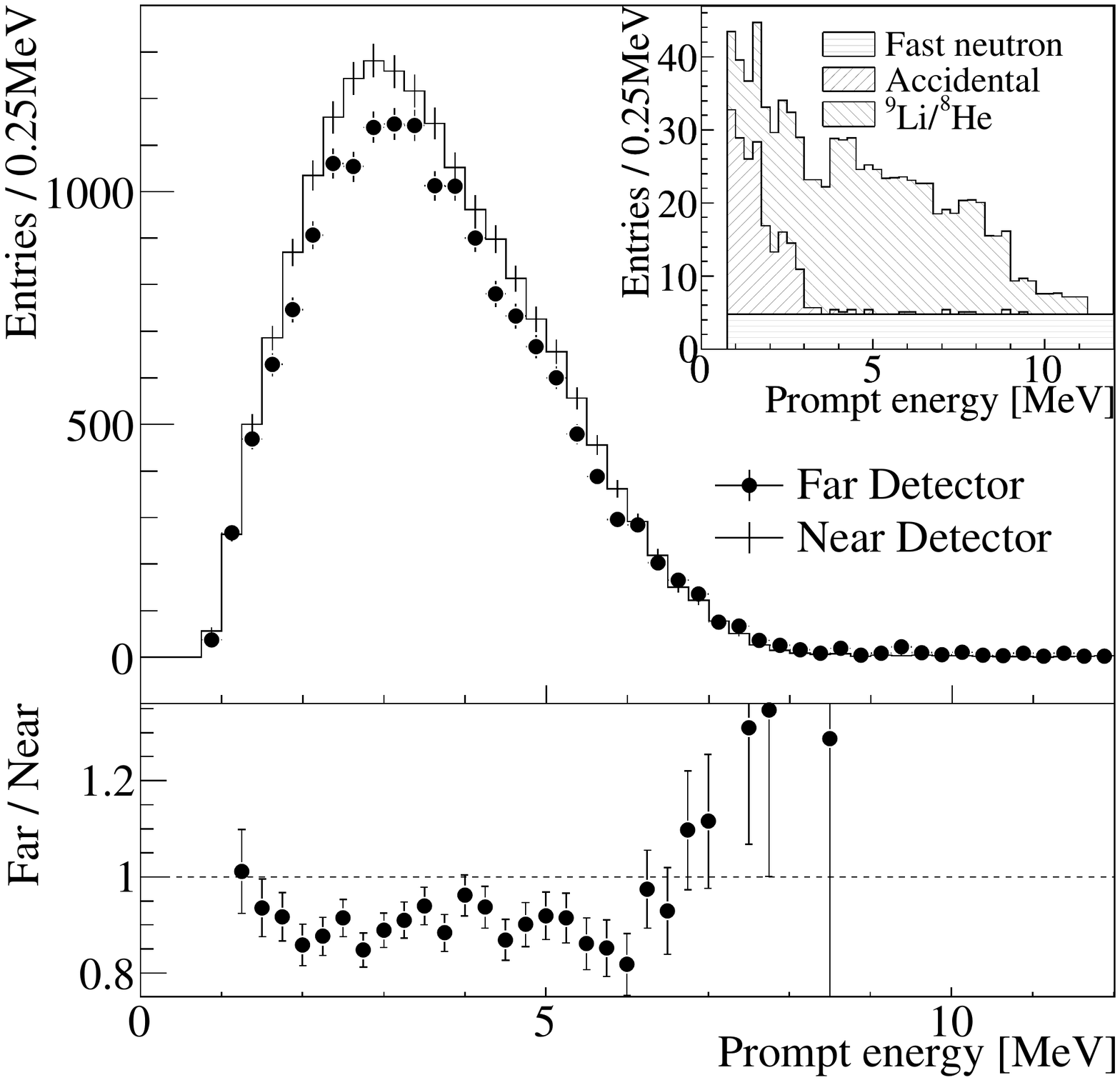}
\caption{Reconstructed energy spectra from Double Chooz (L) and RENO (R) . The insets show the background contributions, the sum of which is shown in green on the Double Chooz histogram, but removed from the main RENO plot.}
\label{reactor-spec}
\end{figure}

\subsection{Results from Double Chooz and RENO}
Results from Double Chooz and RENO both favour a non-zero value of $(\sin^2(2\theta_{13})$.  For the Double Chooz data, the best fit value of $\sin^2(2\theta_{13})$ is \mbox{$0.086\pm0.041\textrm{(stat.)}\pm0.030\textrm{(sys.)}$}, and the hypothesis that $\sin^2(2\theta_{13})=0$ is disfavoured at 94.6\% C.L.  The higher event count of the RENO experiment result in more precise results: \mbox{$\sin^2(2\theta_{13}) = 0.103\pm 0.013\pm0.041\textrm{(stat.)}\pm0.030\textrm{(sys.)}$}, and $\sin^2(2\theta_{13})=0$ is disfavoured at $4.9\sigma$ C.L.   The observed energy spectra and ratios to expectations are shown ing \fig{reactor-spec}. 

\section{Summary}
In the space of ten months five experiments have produced results consistent with a value of \mbox{$\sin^2(2\theta_{13}) \sim 0.1$}.  As a result it can now be considered `well established' that the $U_{e3}$ element of the PMNS matrix is non-zero, with a global level of significance well over the conventional $5\sigma$ discovery threshold.

\begin{figure}
\centering
\includegraphics[width = 0.75\columnwidth]{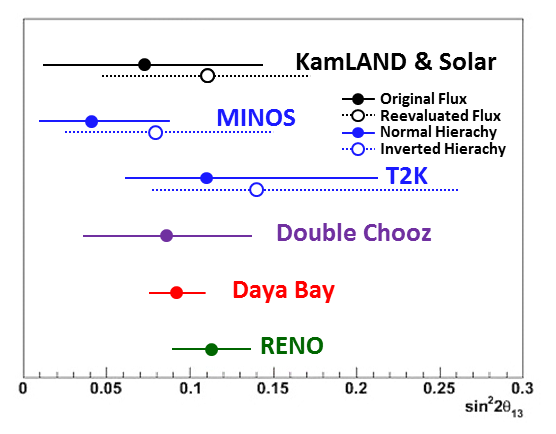}
\caption{Summary of all non-zero measurements of $\sin^2(2\theta_{13})$, as of May 2012.  Figure adapted from presentation by Y.\,Nakajima (Daya Bay)} 
\label{all_results}
\end{figure}

A summary of these measurements, plus the previous hints from global fits, is provided in \fig{all_results}.  With the exception of MINOS, all the experiments described in this talk are expecting to take more data, so the precision on the measured value will almost certainly improve significantly. The differences between the long-baseline and reactor measurements will now come to the fore, with reactor experiments providing precise measures of \modsq{U_{e3}}, while improved measurements from T2K can be used to study CP violation, $\theta_{23}$, and even the mass hierarchy (via matter effects). The field as a whole is therefore rapidly transitioning from $\theta_{13}$ as an unknown parameter, towards $\theta_{13}$ as a well-measured input for investigating other parameters.

\begin{acknowledgments}
Work supported by the Japan Society for the Promotion of Science (JSPS) Postdoctoral Fellowship for Foreign Researchers  
\end{acknowledgments}


\bigskip
\begin{thebibliography}{9}   
\bibitem{tbm}
P.\,F.\,Harrison, D.\,H.\,Perkins and W.\,G.\,Scott, Phys.\,Lett. \textbf{B530} (2002) 167, arXiv:hep-ph/0202074

\bibitem{Freund}
M.\,Freund, Phys.\,Rev. \textbf{D64} (2001) 053003,\\ arXiv:hep-ph/0103300 \\Though not exact, this is also good enough for analyses of current data.

\bibitem{MINOS}
MINOS collaboration, \prl\textbf{107} (2011) 181892, arXiv:1108.0015 [hep-ex]

\bibitem{T2K}
T2K collaboration, \prl\textbf{107} (2011) 041801, arXiv:1106.2822 [hep-ex]

\bibitem{DBatFPCP}
J.\,L.\,Liu (Daya Bay collaboration), These proceedings FPCP2012-32 

\bibitem{DC}
Double Chooz collaboration, \prl\textbf{108} (2012) 131801, arXiv:1112.6353 [hep-ex]

\bibitem{DB}
Daya Bay collaboration, \prl\textbf{108} (2012) 171803, arXiv:1203.1669 [hep-ex]

\bibitem{RENO}
RENO collaboration, \prl\textbf{108} (2012) 191802, arXiv:1204.0626 [hep-ex]

\end{thebibliography}
\end{document}